\documentclass[11pt]{article}
\usepackage{moriond2,epsfig}
\usepackage{graphicx}
\usepackage{floatflt}
\def\be{\begin{equation}}
\def\ee{\end{equation}}
\def\bea{\begin{eqnarray}}
\def\eea{\end{eqnarray}}

\newcommand{\met} {\hbox{E\kern-0.5em\lower-0.1ex\hbox{/}}_T}

\begin{document}
\vspace*{4cm}
\title{W and Z Cross Sections at the Tevatron}

\author{ T. Dorigo \\
(on behalf of the CDF and D0 collaborations)}

\address{Dipartimento di Fisica ``G.Galilei'', Via Marzolo 8,\\
35131 Padova, Italy}

\maketitle\abstracts{
The CDF and D0 experiments at the Tevatron have used 
$p\bar{p}$ collisions at $\sqrt{s}=1.96$ TeV to measure 
the cross section of $W$ and $Z$ boson production using several 
leptonic final states. An indirect measurement of the total $W$ 
width has been extracted, and the lepton charge
asymmetry in Drell-Yan production has been studied up to invariant
masses of $600$ GeV$/c^2$.
}

\section{Introduction}

The Tevatron collider has undergone a massive upgrade
during the last five years. The construction of a new main injector
with a recycler ring, and the improvements done to the antiproton
source and booster ring promise an increase of instantaneous luminosity
by an order of magnitude over Run I. The CDF and D0 experiments have
also recently completed a major upgrade.
CDF was refurbished with an 
entirely new tracking system, with up to 8 silicon layers providing
precise measurement of track parameters close to the beam line, and
with a new calorimeter for intermediate rapidities; the muon system
has been extended to record tracks up to $|\eta|<1.5$. D0
was endowed with a $2T$ axial field, and new silicon and fiber trackers;
mini-drift tubes for forward muons and a new preshower have also been added.

The collider experiments have started recording $p\bar{p}$ interactions
with their full functionality during 2002, and have used the physics-quality
data collected till January 2003 to
measure the production rate of $W$ and $Z$ bosons, which constitute 
the starting point for many high-$P_T$ physics studies at the Tevatron,
and are fundamental ``standard candles'' with which to understand and check
detector performance.

\section{W and Z Cross Sections \label{s:wzxs}}

W and Z bosons are produced at the Tevatron
through $q\bar{q}^{(')}$ annihilation.
The cleanest signatures involve high-$P_T$
electrons or muons: $W \to e \nu$, $\mu \nu$, and $Z \to e^+ e^-$, 
$\mu^+ \mu^-$ all enjoy very small background contaminations.
The $W \to \tau \nu$ decay can also be observed with
clarity although backgrounds are much larger;
other decays are hard to exploit or unobservable.

D0 and CDF presented their first $W$ and $Z$ cross section measurements at 
$\sqrt{s}=1.96$ TeV in the summer 2002~\cite{ICHEP02}. 
Many of those results have now been updated using total integrated 
luminosities of $L=72.0\pm4.3~pb^{-1}$ (CDF) and $L=31.8\pm 3.2~pb^{-1}$ (D0).

\subsection {W Cross Section Measurements}

\begin{floatingfigure}{7cm}
\begin{center}
\includegraphics[width=7cm,height=7cm,clip=]{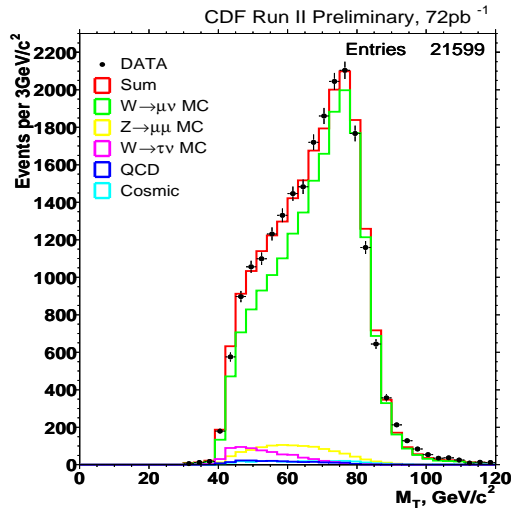}
\caption{\em Transverse mass distribution of $W \to \mu \nu$ candidates
             collected by the CDF II experiment in 
             $72~pb^{-1}$ of Run II data.}
\label{f:wmunu}
\end{center}
\end{floatingfigure}

CDF collects $W \to e \nu$ decays with a trigger selecting
high-$E_T$ central electron candidates; after 
requiring one tight electron with $E_T>25$ GeV matched to a track of 
$P_T>10$ GeV$/c$ and missing transverse energy $\met >25$ GeV, $38628$ 
events are left in the data. The main background source is from $QCD$ 
dijet events where a jet mimics the electron signal and large $\met$
is due to a second poorly measured jet: using events with
non-isolated electron candidates or small $\met$, $1344\pm82\pm672$ events
are estimated from that process. Additional backgrounds from $W \to \tau \nu$ 
decays ($768\pm22$ events) or misidentified $Z \to ee$ decays ($344\pm17$) 
are estimated from Monte Carlo simulations.

The acceptance is $A_{e \nu}=23.4\pm0.05$ (stat.)$\pm0.70$ (syst.)$\%$; 
the systematic error is mainly due to the uncertainty in the 
parton distribution functions (PDF), that affect the fraction of decays 
yielding central electrons by $0.58\%$, and to the knowledge of the 
amount of material in the tracking volume ($0.29\%$).

The result is $\sigma_W B(W \to e \nu)=2.64\pm 0.01$ (stat.) $\pm0.09$ (syst.)
$\pm0.16$ (lum.)$nb$, in good agreement with NNLO 
calculations~\cite{WJStirling}  ($2.731\pm0.002~nb$).

CDF also measures $\sigma_W B(W \to \mu \nu)$. From central high-$P_T$ muon
triggers, events with a clean muon candidate are selected if the muon
has $P_T>20$ GeV$/c$ and if $\met>20$ GeV. The transverse mass spectrum
of the $21,599$ $W$ candidates is shown in Fig.~\ref{f:wmunu}.
Backgrounds in this channel include cosmic rays, QCD processes, and 
misidentified boson decays ($Z \to \mu \mu$, $W \to \tau \nu$);
their sum is estimated at $10.82 \pm 0.18$ (stat.) $\pm 0.96$ (syst.)$\%$.
The total acceptance is $A_{\mu \nu} =14.8\pm0.1$ (stat.) $\pm0.5$ 
(syst.)$\%$; systematics are 
dominated by the uncertainty in the PDF ($0.41\%$)
and by the measurement of the $W$ recoil ($0.23\%$).
The result is $\sigma_W B(\mu \nu)=2.64 \pm 0.02$ (stat.)$ \pm 0.12$ 
(syst.) $\pm 0.16$ (lum.)$nb$, again in good agreement with NNLO
calculations.

$W\to \tau \nu$ candidates can also be selected at CDF
by collecting events with a $\met>25$ GeV hardware trigger at level 1, 
complemented by a subsequent software filter for $\tau$ identification
at trigger level 3.
Monojet candidates are kept if they have a single $E_T>25$ GeV jet 
containing one $P_T>4.5$ GeV$/c$ charged track in a $10^\circ$ cone 
and no other track within $30^\circ$, and if $\met>25$ GeV; tight 
electron events from $W \to e \nu$ decay are explicitly removed.
2345 events pass the selection, with an
estimated background of $612\pm61$ events mainly due to QCD processes.
From those numbers CDF computes 
$\sigma_W B(\tau\nu)=2.62\pm0.07$ (stat.) $\pm 0.21$ (syst.) 
$\pm0.16$ (lum.)$nb$.
Moreover, using the previously excluded $W \to e\nu$ signal in the 
same dataset, it is possible to extract the ratio of coupling constants
$G_\tau/G_e=0.99 \pm 0.04$ (stat.) $\pm 0.07$ (syst.).

\begin{figure}[h!]
\epsfig{file=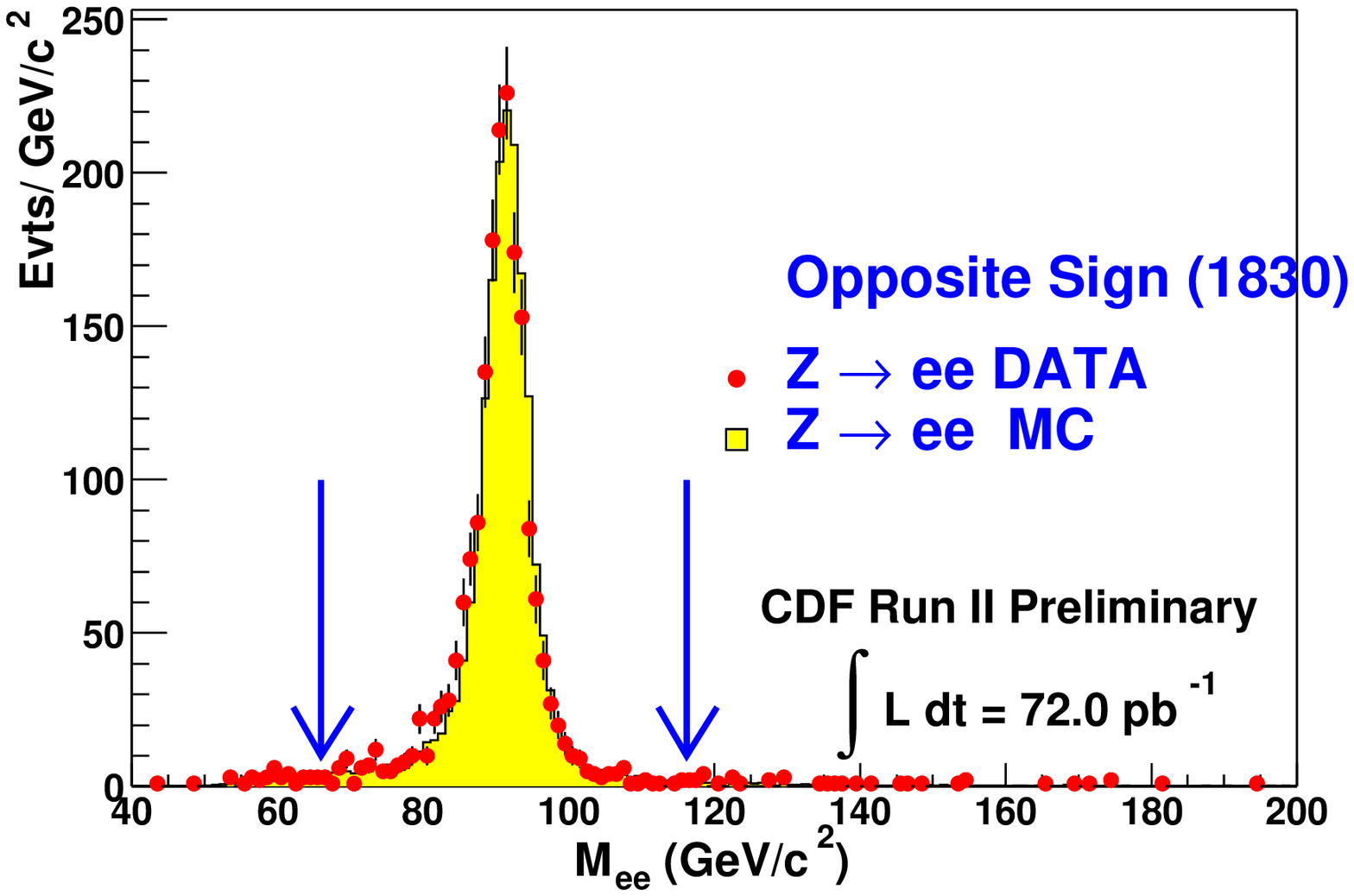, width=8cm, clip=}
\epsfig{file=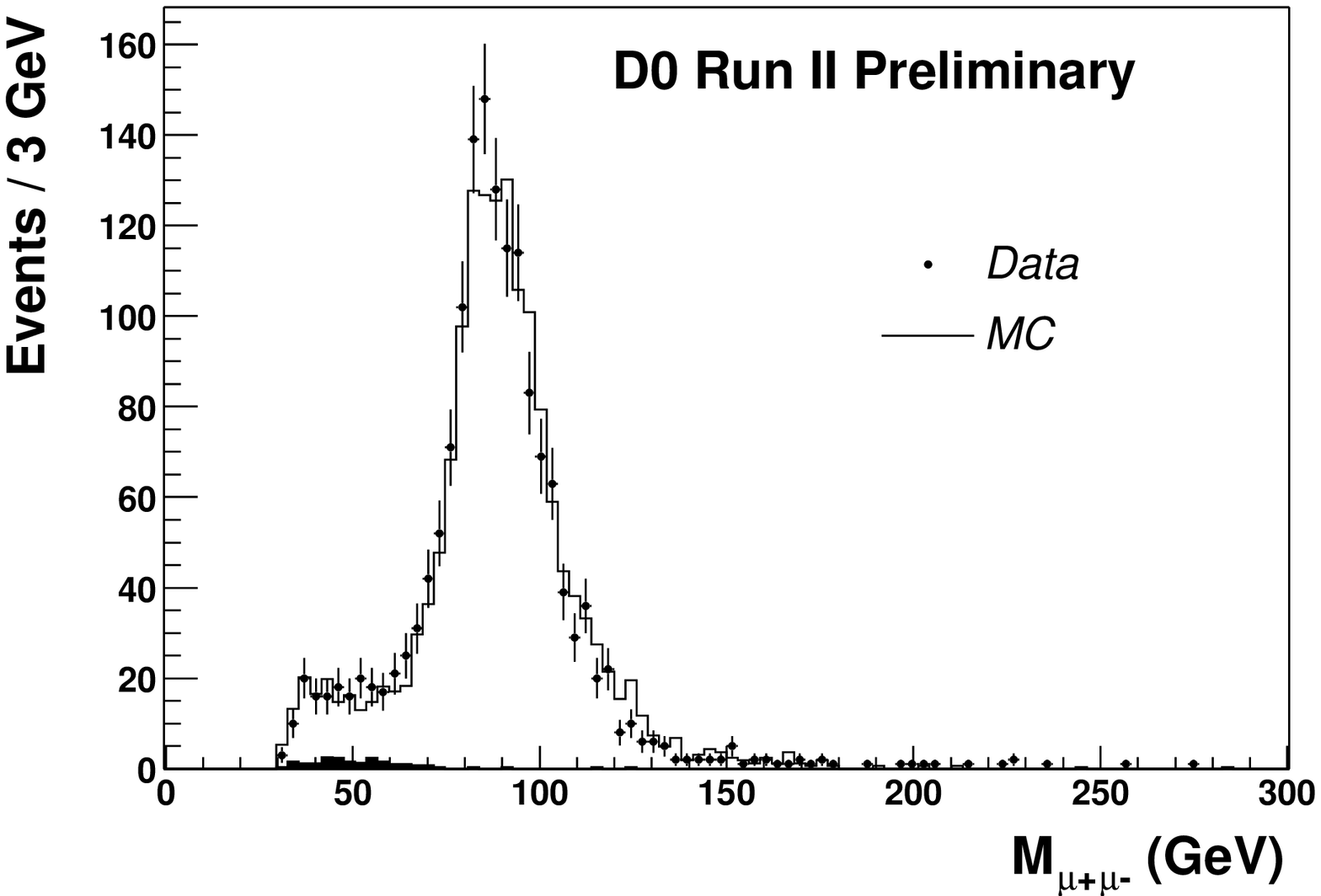,width=7.7cm,clip=}
\caption { Left: invariant mass distribution of electron-positron pairs
           collected by CDF. Right: invariant mass distribution of
           $\mu^+ \mu^-$ pairs collected by D0.}
\label{f:zpeaks}
\end{figure}

\subsection {$Z$ Cross Section Measurements}

To select $Z \to ee$ candidates, CDF requires two central electrons 
with opposite charge, $E_T>25$ GeV, and $P_T>10$ GeV$/c$; they must have
invariant mass in the $66<M_{ee}<116$ GeV$/c^2$ range.
1830 evts are thus collected (see Fig.~\ref{f:zpeaks}, left), 
with $10\pm5$ estimated from background sources.
The total acceptance is $A_{ee}=11.49 \pm0.07$ (stat.) $\pm 0.64$ (syst.)$\%$,
where systematic errors are mostly due to PDF and modeling of tracker material.
The cross section is measured at 
$\sigma_Z B (Z\to ee) = 267.0 \pm 6.3$ (stat.) $\pm 15.2$ (syst.) $\pm 16.0$
(lum.)$pb$, higher but consistent with the NNLO calculation~\cite{WJStirling} 
of $250.5 \pm 3.8~pb$.

\begin{figure}[h!]
\epsfig{file=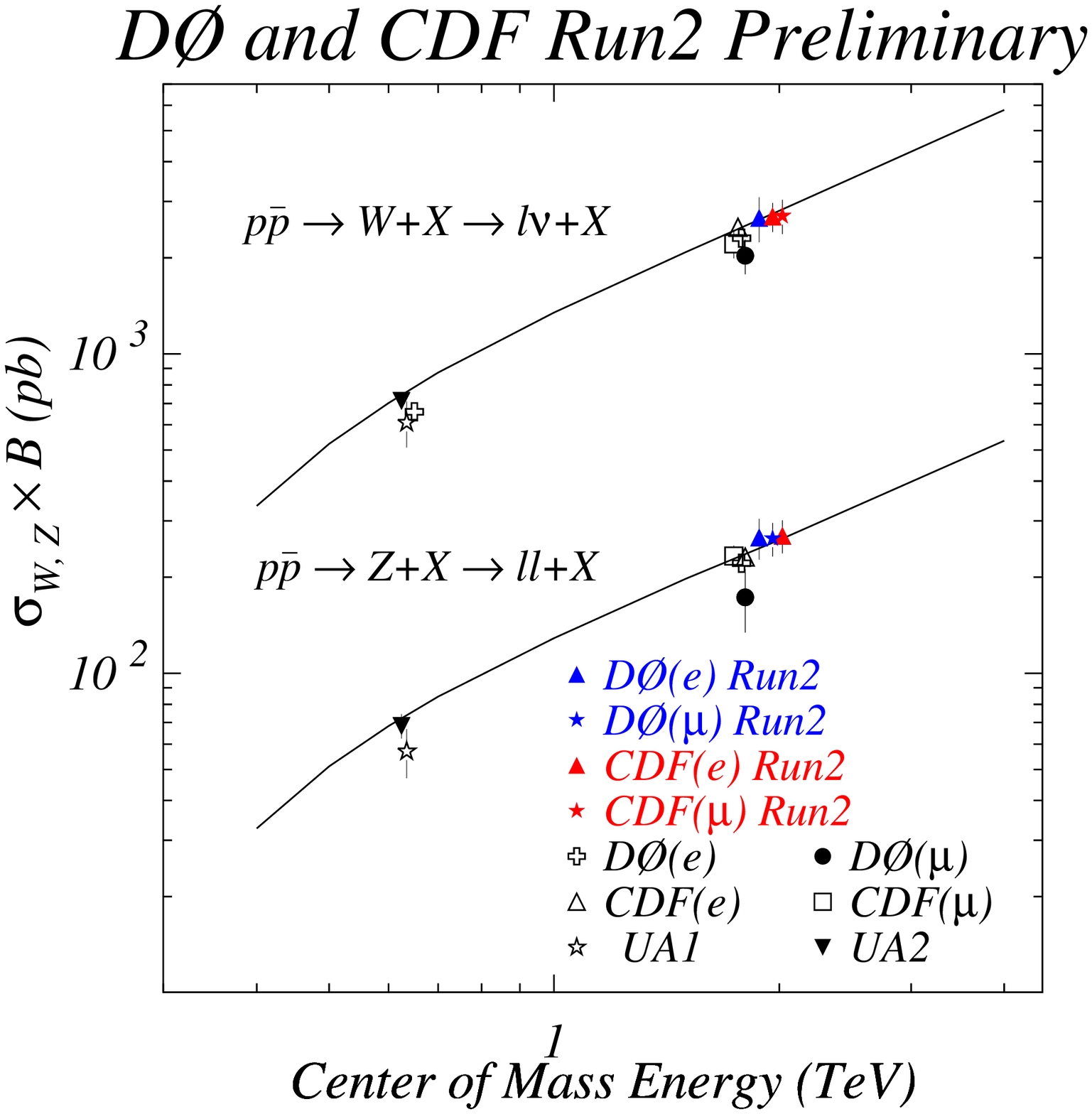,width=8.5cm,clip=}
\epsfig{file=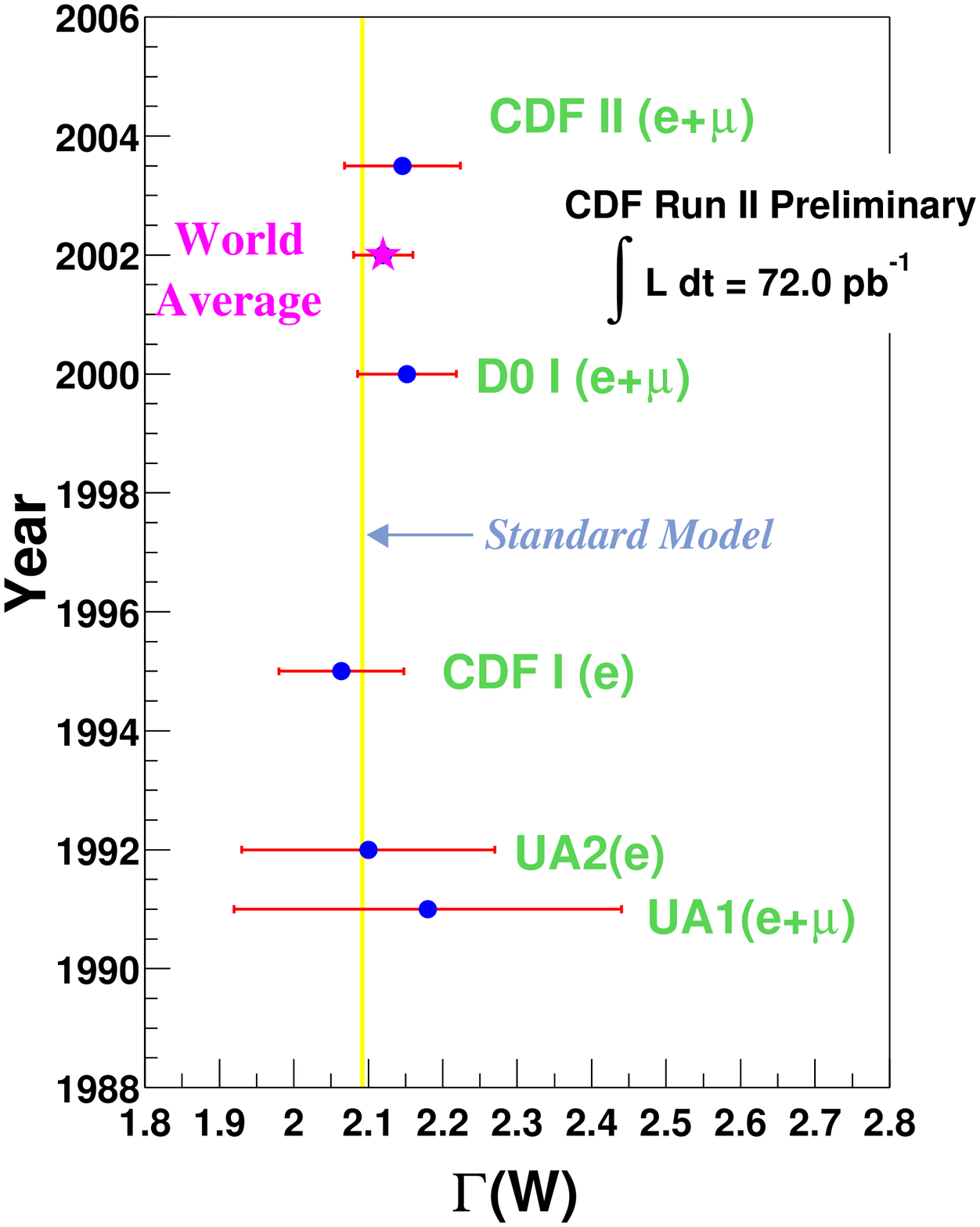,width=7cm,clip=}
\caption{ Left: comparison of cross section measurements for $W$ and $Z$ 
bosons with NNLO calculations (full lines).
Right: comparison of the new CDF result 
on the $W$ boson width with previous determinations.}
\label{f:gamma_w}
\end{figure}

Both CDF and D0 have updated their measurement of $\sigma_Z B(Z \to \mu \mu)$.
From 1632 events containing two muon candidates with $P_T>20$ GeV$/c$, 
CDF measures $\sigma_Z B(\mu \mu)=246\pm6$ (stat.)
 $\pm12$ (syst.) $\pm15$ (lum.)$pb$.
D0 bases the analysis on data collected between September 2002
and January 2003. 1585 events are selected by requiring two muon 
candidates of $P_T>15$ GeV$/c$, separated in $\eta-\phi$ by more than 
$\Delta R_{\mu\mu}>2.0$ (Fig.~\ref{f:zpeaks}, right); 
the estimated background is $1.5\pm1.0\%$. The result is
$\sigma_Z B(\mu \mu)=263.8\pm6.6$ (stat.) $\pm17.3$ (syst.) 
$\pm26.4$ (lum.)$pb$.

\section{ Forward-Backward Dielectron Asymmetry } 

\begin{floatingfigure}{9cm}
\includegraphics[width=7.8cm,clip=]{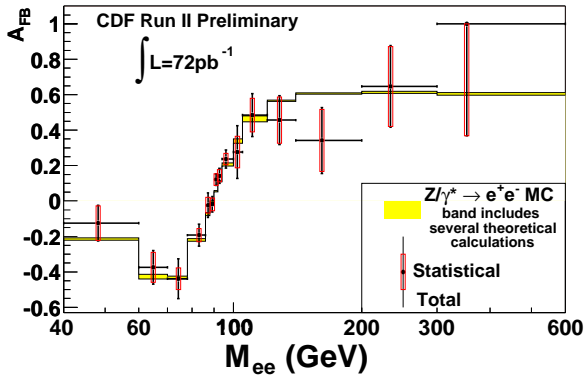}
\caption{ The forward-backward asymmetry of elec\-tron-positron pairs measured
          by CDF, compared to theory predictions. }
\end{floatingfigure}

\noindent
CDF uses 5438 
events with $40<M_{ee}<600$ GeV$/c^2$ from 
the $p\bar{p} \to Z(\gamma^*)X \to eeX$ process
to measure the forward-backward charge asymmetry $A_{FB}$ over
a wide range of $Q^2$.
Backgrounds are due to QCD processes and are estimated from 
same-sign or non-isolated elec\-tron-positron pairs; they amount to 
$N_{QCD}^{cc}=21.5\pm5.9$ events for central-central  
pairs, and to $N_{QCD}^{cp}=128\pm65$ events for events with one forward 
($|\eta|<2$) electron.
The new $A_{FB}$ measurement is shown in Fig.~4; 
results agree with theoretical predictions.
\vspace{.5cm}

\section {$W/Z$ Ratio and $W$ Width}

Using the cross section measurements in the electron channel
presented in Sec.~\ref{s:wzxs},
CDF computes the ratio 
$R_e = \sigma_W B (W \to e \nu) / \sigma_Z B (Z \to ee) =$
$9.88 \pm 0.24$ (stat.)$ \pm 0.47$ (syst.),
lower but consistent with NNLO predictions 
($R_{th}=10.66\pm0.05$)~\cite{WJStirling}.
Using the LEP measurement of $B(Z\to ee)$~\cite{PDG2002} plus
theoretical predictions for $\sigma_W/\sigma_Z$ and 
$B(W\to e\nu)$~\cite{WJStirling}, the measurement of $R_e$ can be converted 
into an indirect determination of 
$\Gamma_W = 2.29 \pm 0.06$ (stat.) $\pm  0.10$ (syst.) GeV, in good
agreement to the most recent average of $2.118 \pm 0.042$ GeV~\cite{PDG2002}.

In addition, using their muon decay results (Sec.~\ref{s:wzxs}), CDF finds 
$R_\mu=10.69\pm0.27$ (stat.) $\pm0.33$ (syst.) and 
$\Gamma_W=2.11\pm0.05$ (stat.) $\pm0.07$ (syst.) 
GeV. The electron and muon channel results can be combined into
$\Gamma_W=2146\pm78$ MeV.
Fig.~\ref{f:gamma_w} compares that determination to previous
measurements of the $W$ boson width.

\section { Concluding Remarks}

$W$ and $Z$ cross sections in $p\bar{p}$ collisions 
have been shown to rise with $s$ in accordance
with NNLO calculations. The Drell-Yan asymmetry
has been probed up to the invariant mass $M_{ee}=600$ GeV$/c^2$.
From the cross section measurements an indirect determination
of the total width of the $W$ boson has been extracted, in good agreement
with Standard Model predictions. 
In conclusion, Run II at the Tevatron is starting to deliver: although
the CDF and D0 detectors have not yet been exploited to their fullest
potential, the results discussed here have already reached accuracy 
levels similar to Run I ones. 


\end{document}